\documentclass[aps,prb,twocolumn]{revtex4}
\usepackage{graphics,amsmath}
\begin{document}
\title{Electronic susceptibilities in systems with anisotropic Fermi surfaces}

\author{S. Fratini}
\altaffiliation{Permanent address, 
LEPES/CNRS, BP 166, F-38042 Grenoble Cedex 9, France}
\author{F. Guinea}
\affiliation{Instituto de Ciencia de Materiales de Madrid,
CSIC, Cantoblanco, E-28049 Madrid, Spain.}

\begin{abstract}
The low temperature dependence of the spin and charge susceptibilities
of an anisotropic electron system in two dimensions is
analyzed. It is shown that the presence of inflection points
at the Fermi surface leads, generically, to a 
$ T  \log T $ 
dependence, and a more singular behavior,
$\chi \sim  T ^{3/4} \log T$, is also possible. Applications to quasi
two-dimensional materials are discussed.  
\end{abstract}

\date{\today}
\maketitle

\section{Introduction}
The possible existence of quantum critical points in the phase diagrams
of many materials has led to a detailed study of
the low temperature behavior of the susceptibilities of
electron systems. The critical properties of the system are
determined by the energy and momentum dependence of the response function
of the electron system associated to the order
parameter in the ordered phase\cite{H76,M85,M93}.
It has been shown that the low temperature spin susceptibility
of the isotropic electron liquid has an unexpected non analytic
dependence on temperature, when high order perturbative corrections
are considered\cite{BKV97}. These corrections are irrelevant
in the Renormalization Group sense\cite{S94,P92,MCC98}, and do not
modify the basic properties of the electron liquid, as described by
Landau's theory. However, they can lead to unexpected power
law dependences in many physical quantities at low temperatures,
or change the order of the phase transitions\cite{BKV99}. 
The origin of these non analiticities in
homogeneous response functions has been traced back to the 
special properties of $2 k_F$ scattering in the isotropic
electron liquid\cite{CM01}.

It is well known that anisotropic Fermi surfaces can have regions
where scattering becomes more singular than in the isotropic
electron liquid, the so called `` hot spots ". When two
portions of the Fermi surface are flat and
parallel, nesting occurs, and
the susceptibilities diverge logarithmically,
${\rm Re} \chi ( {\bf \vec{Q}} , \omega ) \propto \log ( \Lambda / \omega )$,
where ${\bf \vec{Q}}$ is the nesting vector. A saddle point
in the density of states leads also to logarithmic divergences
in two dimensions. The hot spots at the Fermi surface can be
characterized by the frequency dependence of ${\rm Im}
\chi ( {\bf \vec{Q}} , \omega )$, where
${\bf \vec{Q}}$ spans the hot spots. The usual behavior in
a Fermi liquid is ${\rm Im} \chi ( {\bf \vec{Q}} , \omega )
\propto | \omega |$, in any dimension $D$. For an isotropic
Fermi surface, if $| {\bf \vec{Q}} | = 2 k_F$, one has
${\rm Im} \chi ( {\bf \vec{Q}} , \omega )
\propto | \omega |^{(D-1)/2}$. For $D=1$ the imaginary part
of the $2 k_F$ susceptibility approaches a constant at low
frequencies. By a Kramers Kronig transformation, it can be
shown that the real part should diverge logarithmically,
leading to the deviations from Landau's theory which
characterize a Luttinger liquid.    

\begin{figure}
\resizebox{7cm}{!}{\rotatebox{0}{\includegraphics{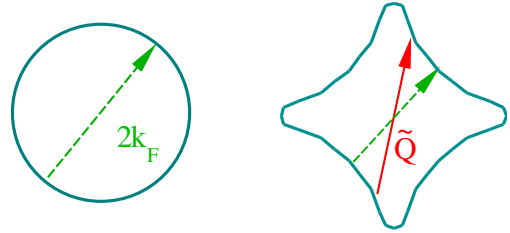}}}
\caption{Examples of Fermi surfaces in 2 spatial dimensions. Left: circular,
  all wavevectors of modulus $2k_F$ are sources of enhanced
  scattering. Right: anisotropic,
  wavevectors such as $\tilde{\mathbf{Q}}$ 
  connecting 2 inflection points give rise to
  anomalous scattering (continuous arrow), while the rest of the FS
  gives rise to a behavior similar to the isotropic case (dashed arrow).}
\label{fig:FS-ex}
\end{figure}

It is also possible to show that, when ${\bf \vec{Q}}$ connects
two saddle points in an anisotropic Fermi surface,
${\rm Im} \chi ( {\bf \vec{Q}} , \omega )
\propto | \omega |^{(D-2)/2}$. This result implies the existence
of logarithmic divergences for $D=2$, which have been extensively studied
in relation to high T$_{\rm c}$ superconductors\cite{VH}, and lead
to deviations from Landau's theory\cite{GGV96}. In addition to
saddle points, a  generic anisotropic Fermi surface can show 
inflection points (see Fig.\ref{fig:FS-ex}). The existence of these
points at the Fermi surface, which  do not require any
special fine tuning of the chemical potential,
leads to\cite{GGV97} ${\rm Im} \chi ( {\bf \vec{Q}} , \omega )
\propto | \omega |^{(D-2)/2 + 1/4}$. For $D = 2$, scattering between
these points is more singular than the $2 k_F$ scattering 
considered previously.

In the present work, we analyze scattering at inflection points
in a two dimensional anisotropic Fermi surface. In the next section,
we present the main features of the two loop calculation,
extending the method used in reference\cite{CM01}. The main results are
obtained in section III, while the finer details of the calculation
are deferred to the appendices. Applications to Fermi surfaces of
different shapes are given in section IV, and section V discusses
the main results of our work.

\section{The method} 
We consider a system of two-dimensional (2D) fermions interacting
through a generic short-ranged effective potential $U(q)$. For the  sake of 
simplicity, we shall also assume that the interaction only 
affects electrons of opposite spins, which is a reasonable
approximation when the momentum dependence of $U(q)$ is weak.
It was shown in refs.\cite{BKV97,CM01}
 that while the lowest order ($\propto U$) perturbative corrections
are well behaved, higher order corrections can 
lead to an anomalous behaviour in the low-energy properties of 
the Fermi liquid. To be more precise, the uniform spin susceptibility 
of a 2D electron system shows a linear $T$ dependence, 
which contradicts the usual Sommerfeld expansion in powers of
$(T/E_F)^2$.  Such anomalous behaviour was traced back to the 
peculiarities of $2k_F$ scattering, i.e. the occurrence of
particle-hole pairs lying on opposite sides of the Fermi surface (FS). 
This special wavevector plays a key role in the $q$-dependent susceptibility 
of electronic systems already in the non-interacting case, 
with the appearance of a square-root singularity around $2k_F$ 
which is directly related to the jump in the occupation number.
If one considers the \textit{uniform}
susceptibility, though,  the singularities associated with $2k_F$
scattering can only show up indirectly through the excitation of a
\textit{virtual} particle-hole pair, which explains  
the absence of anomalous corrections at lowest order in the interaction
strength.
%is not surprising:  no particle-hole pairs
%can be excited at $2k_F$.

\begin{figure}
\resizebox{8.5cm}{!}{\rotatebox{0}{\includegraphics{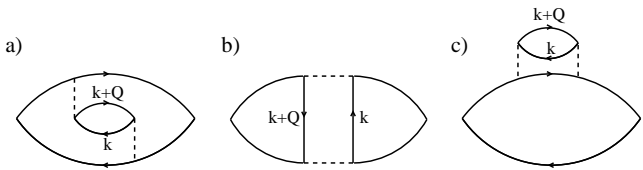}}}
\caption{The second order diagrams yielding the leading temperature
  dependence of the susceptibility.}
\label{fig:susc}
\end{figure}

In the general (non circular) case, among all the
wavevectors connecting opposite sides of the  FS, the
inflection points  play a special role, due to the flatness of
the Fermi surface (the extreme case being the one of a perfectly flat 
FS, or perfect nesting, which leads to strong instabilities).
According to the previous discussion,  the
second-order diagrams which lead to non-standard behavior are the 
ones containing a particle-hole bubble whose transferred momentum can match 
the special value $\tilde{\mathbf{Q}}$. Such diagrams are depicted in 
figure \ref{fig:susc}.
%The three diagrams all have the same $T$-dependence.
Diagrams a) and b) are vertex corrections.
They have opposite signs and 
cancel in the case of a perfectly
$q$-independent interaction potential: the fermion propagators
involved are the same in both diagrams, the only difference being in 
the momentum carried by the interaction. To be specific, 
 with the notations of 
figure \ref{fig:susc}, diagram a) is proportional to 
$U(\tilde{\mathbf{Q}})^2$, while diagram b) involves some momentum
average of the interaction, and there is no reason for a perfect
cancellation in the general case. 
Diagram c) is a self-energy correction, and will be considered
separately.
% at the end of the section.
%In the following section, we shall give the full details of the
%calculation of the vertex diagrams with various possible shapes of the FS, and
%eventually give a hint of how to calculate the self-energy diagram. 

%In the next section, we shall proceed to the evaluation of the leading 
%temperature dependence of such vertex diagrams.

\section{Uniform susceptibilities}

After integration over Matsubara frequencies, the vertex correction a) 
of figure \ref{fig:susc} in the zero-frequency, zero-momentum limit reads:
% a factor of two if sum over different spins
%\begin{widetext}
\begin{equation}
\chi(T) =\int\! \frac{d^2p}{(2\pi)^2}  \frac{d^2Q}{(2\pi)^2} U(\mathbf{Q})^2
\Delta(\xi_{\mathbf{p}})\Delta(\xi_{\mathbf{p+Q}})
L(\xi_{\mathbf{p+Q}}-\xi_{\mathbf{p}},\mathbf{Q})
\label{eq:susc}
\end{equation}
%\end{widetext}
where the Lindhard function in 2 space dimensions is defined as
\begin{equation}
  L(i\Omega,Q)=  \int  \frac{d^2k}{(2\pi)^2}\frac{n(\xi_\mathbf{k})-n(\xi_\mathbf{k+Q})}
{i\Omega+\xi_\mathbf{k}-\xi_\mathbf{k+Q}} 
\label{eq:Lind}
\end{equation}
and
\footnote{This is the finite
  temperature generalization, within our perturbative treatment,  
  of the well-known Fermi Liquid formula expressing the singularity of 
  $G(k)G(k+q)$ lines in the vertex function, see e.g. A.A. Abrikosov,
  L.P. Gorkov
  and I.E. Dzyaloshinski, \textit{Methods of Quantum Field Theory in
  Statistical Physics} (Dover, New York, 1975).} 
$\Delta(\xi)=\beta/4\cosh^2(\beta\xi/2)$. 
Such  $\Delta$-functions, which constrain momenta to lie
within a shell of thickness $\sim T$ from the Fermi surface, 
are strongly temperature dependent, and they 
are responsible for the leading 
temperature dependence of the susceptibility. Indeed, in
(\ref{eq:susc}) we have omitted  
terms proportional to $\Delta(\xi_{\mathbf{p+Q}}) n(\xi_{\mathbf{p}})$ 
coming from the  poles of the Lindhard function,  
which are less $T$-dependent since they receive contributions mainly
from regions far from
the Fermi surface. 
 Taking advantage of 
 time-reversal symmetry ($\xi_\mathbf{k}=\xi_{-\mathbf{k}}$), we can write
\begin{equation}
  L(\Delta\xi_p,Q)=\int \frac{d^2k}{(2\pi)^2}n(\xi_\mathbf{k})\left\lbrack
\frac{1}{\Delta\xi_\mathbf{p}-\Delta\xi_\mathbf{k}}- 
\frac{1}{\Delta\xi_\mathbf{p}+\Delta\xi_\mathbf{k}} \right\rbrack
\label{eq:Lind-sep}
\end{equation}
where we have defined  $\Delta\xi_{\mathbf{k}}=
\xi_{\mathbf{k+Q}}-\xi_{\mathbf{k}}$. 
%In order to calculate the leading temperature dependence in $d=2$, one
%essentially has to evaluate the momentum space available in the $6$
%momentum integrations, taking account of the constraints imposed by
%the $\Delta$ functions. 
As was pointed out in the previous section, the most
singular contributions to (\ref{eq:susc}) 
come from regions where the momentum $\mathbf{Q}$
flowing through the Lindhard function $L$ connects parts of the FS which are 
almost parallel, since this makes the denominators in
eq. (\ref{eq:Lind-sep}) small on large regions of $k$-space. Otherwise
stated, the scattering processes taking place within a
particle-hole pair are enhanced around special wavevectors
$\tilde{\mathbf{Q}}$ due to the peculiar geometry of the FS.
In the case of a spherically symmetric FS, any
momentum $\tilde{\mathbf{Q}}$ of modulus $2k_F$ is a source of enhanced
scattering, but the deviation from parallelicity is \textit{quadratic} as we
move in the direction tangent to the surface (see
fig. \ref{fig:FS-ex}, left). 
More singular is the case of inflection points occurring when 
the curvature of the FS vanishes, leading to a \textit{cubic}, 
or even \textit{quartic} dispersion (see figure \ref{fig:FS-ex},
right), 
which is 
a quite generic phenomenon when dealing with electrons on a lattice.

In the next subsections, we shall present the calculation of the leading
$T$-dependence of the diagram a) in the simplest
circular case as well as for more complex FS shapes.
%, we shall only quote the
%results, deferring the details of the calculations to the appendices.
The result for diagram b) can obtained by replacing 
$ U(\mathbf{Q})\rightarrow  U(\mathbf{p-k})$ in
eq. (\ref{eq:susc}). This can only lead to a change in 
the prefactors, but will not alter the leading temperature dependence
of the susceptibility. The self-energy diagram c) has a different
structure, and will be analysed at the end of the section.

\subsection{Isotropic Fermi surface}

By choosing an appropriate coordinate system, the 
dispersion relation around any point  on a
spherical FS (and, generically, about non-special points of an
anisotropic FS) can be expanded as:
\begin{equation}
  \xi_{\mathbf{k}}/v=k_y+ak_x^2
\end{equation}
$v$ being the Fermi velocity at that particular point (that we shall
identify as $\tilde{\mathbf{Q}}/2$), and $a>0$ being
related to the FS curvature. The above expression is assumed to be
valid up to a momentum cutoff $\Lambda$ 
which is larger than the one imposed by the
finite temperature $\Delta$-functions \footnote{The momentum cutoff
  should in principle be different in the $x$ and $y$ directions, 
  but this is of no practical relevance here.}. 
Introducing $\mathbf{q}=\mathbf{Q}-\tilde{\mathbf{Q}}$, we can write
\begin{equation}
  \xi_{\mathbf{k+Q}}/v
=-(k_y+q_y)+a(k_x+q_x)^2
\end{equation}
%= \xi_{\mathbf{-k-q}}/v
were we have used the property $\xi_{\mathbf{k+\tilde{Q}}}=\xi_{\mathbf{-k}}$ 
(reflection symmetry). We now change variables to
\begin{eqnarray}
  k_\perp&=&k_y+ak_x^2\label{eq:chvar} \\
  k_\parallel&=&k_x+q_x/2 \nonumber
\end{eqnarray}
such that $\xi_{\mathbf{k}}= vk_\perp$ and
\begin{equation}
 \xi_{\mathbf{k+Q}}/v=-k_\perp-q_\perp+2ak_\parallel^2 +(3/2)aq_\parallel^2 
\label{eq:en-shift}
\end{equation}
Apart from the constant shift $q_x/2$ introduced for later convenience, 
this new coordinate system is locally equivalent to polar
coordinates, which would be the natural choice when dealing with a
perfectly symmetric FS. 
We shall focus on the first term in (\ref{eq:Lind-sep}), which is
independent of  $q_\perp$, and therefore turns out to be the most singular. 
Omitting unimportant
multiplicative factors, we have for the real part of $L$:
\begin{eqnarray}
  L&=&\frac{-1}{v}\int dk_\perp n(vk_\perp) \;
  \int_{-\Lambda}^{\Lambda} dk_\parallel
  \frac{1}{A-2k_\perp+Bk_\parallel^2} \nonumber\\
   &=& \frac{-1}{v\sqrt{B}} \int dk_\perp n(vk_\perp)  
\frac{\theta(A-2k_\perp)}{\sqrt{A-2k_\perp}}
\label{eq:Lind-parab}
\end{eqnarray}
with  $A=2p_\perp-2ap_\parallel^2$ and $B=2a$ (the $\theta$-function
ensures that the integral is real). In the second term of
eq. (\ref{eq:Lind-parab}) we have performed the $k_\parallel$
integration by pushing the momentum cutoff to infinity. The main point 
is that the former expression can now be integrated by parts to give a further
$\Delta$ constraint on $k_\perp$ \cite{CM01}:
\begin{equation}
L=
\frac{1}{\sqrt{B}}\int^{A/2}_{-\Lambda} dk_\perp \sqrt{A-2k_\perp} \Delta(vk_\perp)
+\ldots
\label{eq:part-parab}
\end{equation}
where the ellipsis stands for terms which are not confined to the
region near the FS. By inspection of the results for $A\gg T/v$, $A\ll
-T/v$ and $A\approx 0$ respectively, we see that the
$\Delta(vk_\perp)$ function behaves qualitatively as a
$\delta(vk_\perp+T)$. Therefore, to study the temperature dependence of the
susceptibility we can replace the previous expression by
\begin{equation}
L \sim \frac{\sqrt{A/2+T/v}}{v\sqrt{a}}
\label{eq:part-parab-result}
\end{equation}
where there is an implicit 
$\theta$-function of the argument of the square root. We are left
with a tractable expression for the real part of the Lindhard
function, that we shall use to evaluate the 2-loop diagram of figure
\ref{fig:susc}.a.

We now perform the remaining integrals in (\ref{eq:susc}) 
in the following order:
$dq_\perp,dp_\perp$ then $dp_\parallel$ and $dq_\parallel$. The
 first integral is trivial, since $q_\perp$ only enters in
 $\Delta(\xi_{\mathbf{p+Q}})$. Moreover, $\xi_\mathbf{p+Q}$ is linear 
in $q_\perp$ (cf. eq. (\ref{eq:en-shift})) so that the integration just
gives $1/v$. The $p_\perp$ integral can also be performed
straightforwardly, and we are left with an expression of the form
\begin{eqnarray}
  \chi &\sim & \frac{\tilde{U}^2}{v^3\sqrt{a}} \int dq_\parallel\int dp_\parallel 
\sqrt{T/v-ap_\parallel^2}  \\
&\sim&  \frac{\Lambda\tilde{U}^2}{v^3\sqrt{a}} \int dp_\parallel 
\sqrt{T/v-ap_\parallel^2} 
\sim \frac{\Lambda\tilde{U}^2}{v^4a} T \nonumber
\end{eqnarray}
where again we have neglected unimportant multiplicative
factors and we have defined $\tilde{U}=U(\tilde{\mathbf{Q}})$. 
Within our treatment, we have recovered the result that 
the spin susceptibility of an isotropic 2D Fermi
liquid is intrinsically 
linear in temperature\cite{CM01}. For a circular FS, this 
can be written as
%in units of the density of states $\nu_0$ at the Fermi level as
\begin{equation}
  \chi(T) =  \chi_0+ \chi_1  T 
\label{eq:chi-parabolic}
\end{equation}
Incidentally, our calculation suggests that the low-temperature 
correction to the susceptibility is positive, in agreement with refs.
\cite{M99,HT98,CM01}.

%the general thermodynamic arguments of ref.\cite{MisawaJPSJ99}
% (see \cite{MisawaJPSJ99} and references therein, where this
%fact is used to explain  the phenomenon of thermally induced
%ferromagnetism\cite{thermferro}).  
%This was also confirmed by numerical\cite{HirashimaJPSJ98} and
%analytical\cite{chitov}  
%evaluation of the perturbative expansion to second order in the
%interaction. 

\subsection{Anisotropic FS with inflection points}

In the vicinity of an inflection point, the dispersion relation 
can be written as
\begin{equation}
  \xi_k/v=k_y-b k_x^3+g k_x^4
\label{dispersion}
\end{equation}
where $b$ and $g$ can be chosen to be positive. A change
of variables similar to eq. (\ref{eq:chvar}) of the previous section leads to
\begin{equation}
  L=\frac{-1}{v}\int dk_\perp n(vk_\perp)\; \int dk_\parallel 
\frac{1}{A-2k_\perp+Bk_\parallel^2+Ck_\parallel^4}
\label{eq:Lind-infl}
\end{equation}
with $A=2p_\perp-3bq_\parallel p_\parallel^2-2gp_\parallel^4$, 
$B=3bq_\parallel$ and $C=2g$.
The $k_\parallel$ integral can be rewritten in the form
\begin{equation}
  I=\frac{1}{C^{1/4}(A-2k_\perp)^{3/4}}\int \frac{dx}{1+\alpha
    x^2+x^4}
\label{eq:I-integral}
\end{equation}
with $\alpha=(B/\sqrt{C})/\sqrt{A-2k_\perp}$. We shall be interested in the 
region close to the edge ($k_\perp\simeq A/2$), where $\alpha$ is large and 
positive. We can then drop the  quadratic term and perform the
integration:
\begin{equation}
  I=\frac{1}
{[bq_\parallel(A-2k_\perp)]^{1/2}}
\label{eq:sqrt-edge}
\end{equation}
This can be integrated by parts in $dk_\perp$ to give
%-\int dk_\perp
%\frac{n(vk_\perp)}{v\sqrt{bq_\parallel}\sqrt{A-2k_\perp}}=
\begin{equation}
L=\frac{1}{\sqrt{bq_\parallel}}
\int^{A/2}_{A/2-b^2q_\parallel^2/g} dk_\perp \sqrt{A-2k_\perp}
\Delta(vk_\perp)
+\ldots
\label{eq:parts}
\end{equation}
where the ellipsis stands for a term which is not confined close to
the FS (the limits of integration account for the condition
$\alpha\gtrsim 1$). 
Provided that the $\Delta$ function lies entirely inside the domain
  of integration, i.e. 
\begin{equation}
q_\parallel>q_{min}=\sqrt{\frac{gT}{b^2v}}
\label{eq:cond}
\end{equation}
the result takes the form
\begin{equation}
L\sim \frac{\sqrt{A/2+T/v}}{v\sqrt{bq_\parallel}}
\end{equation}
%The conditions leading to the above result 
%%Equations (\ref{eq:parts}) and (\ref{eq:cond}) 
%can be interpreted in the following way: the
%validity of the approximation  (\ref{eq:sqrt-edge}) for the edge 
%singularity is set by $q_\parallel$ --- cf. the limits of integration 
%in eq.  (\ref{eq:parts}) --- so that a large $q_\parallel$ is needed 
%in order to get a sizeable contribution from such square-root edge 
%--- condition (\ref{eq:cond}). 
The  region of phase space we have just identified
is the one which gives the leading temperature dependence in the
susceptibility. Indeed, for $k_\perp$ outside the range of integration
considered above (implying $\alpha \lesssim 1$),  the result of the
integral (\ref{eq:I-integral}) is
$I\sim (A-2k_\perp)^{-3/4}$  instead of eq. (\ref{eq:sqrt-edge}),
leading to a weaker (linear) temperature dependence in the final result. The
same holds if we consider a negative $q_\parallel$ ($\alpha<0$).

The calculation now follows the same lines as
in the previous case. The $q_\perp$ integration yields a factor
$1/v$, and the $p_\perp$ integration can be performed by replacing 
$\Delta(vp_\perp)\sim \delta(p_\perp- T/v)/v$, which gives
\begin{eqnarray}
  \chi&\sim &\tilde{U}^2 \int_{q_{min}}^\Lambda dq_\parallel 
\int_{-\Lambda}^\Lambda dp_\parallel 
\frac{\sqrt{T/v-bq_\parallel
    p_\parallel^2-gp_\parallel^4}}{v^3\sqrt{bq_\parallel}}\nonumber \\
&\sim&  \frac{\tilde{U}^2T}{bv^4} \int_{q_{min}}^\Lambda 
\frac{dq_\parallel}{q_\parallel} = 
-\frac{\tilde{U}^2}{bv^4}T \log \left(\frac{gT}{vb^2\Lambda^2}\right)
\label{eq:chi-infl-interm}
\end{eqnarray}
% &=&  \frac{U^2}{v^3} \int_{q_{min}}^\Lambda 
%\frac{dq_\parallel}{\sqrt{bq_\parallel}} 
%\int_{-\Lambda}^\Lambda 
% dp_\parallel  \sqrt{T/v-bq_\parallel p_\parallel^2} \\
%(in the second equation, 
%we have been using the condition $q_\parallel>q_{min}$)
Taking into account the scattering from the regions of the
FS far from the inflection points, whose behavior is given by
eq. (\ref{eq:chi-parabolic}),  
the susceptibility reads
\begin{equation}
  \chi(T) = \chi_0 + \chi_1 T - \chi_1^\prime T \log T
\label{eq:chi-infl}
\end{equation}
Once again, the sign of the correction is such that the
susceptibility increases with temperature. However, the contribution
coming from the other diagram b) has opposite sign. As
a rule of thumb, one can argue that the overall vertex-correction is positive 
if the effective interaction is peaked around $\tilde{\mathbf{Q}}$ and
negative otherwise (it vanishes when the momentum dependence of $U(q)$
is flat, since in that case the two diagrams perfectly cancel).

\subsection{Special inflection points}

\begin{figure}[ht!]
\resizebox{8cm}{!}{\rotatebox{0}{\includegraphics{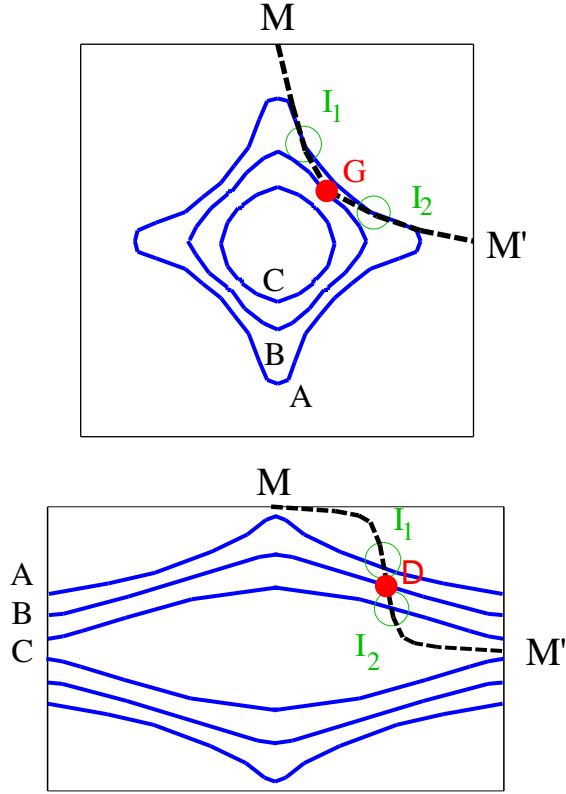}}}
\caption{Top: Fermi surfaces for different fillings in an
anisotropic 2D system with tetragonal symmetry. Curve A has eight
inflection points (only two are shown, I$_1$ and I$_2$). The set of
all these points
define a curve which goes from point M to point M' of the Brillouin Zone
(broken curve). The inflection points merge in pairs when the
Fermi surface is given by curve B. At point $G$ in curve B, 
the cubic term in the dispersion relation parallel to the Fermi surface
vanishes. The Fermi surface labelled C has no inflection points.\\
Bottom: Fermi surfaces for different fillings in an
anisotropic 2D system with orthorhombic symmetry. Curves A and C correspond
to Fermi surfaces with four inflection points. Only one such point
in each curve is shown, I$_1$ and I$_2$. The set of
all these points
define a curve which goes from point M to point M' of the Brillouin Zone
(broken curve). The quartic term in the dispersion relation parallel 
to the Fermi surface changes sign when going from M to M'. Thus, there
is a Fermi surface, schematically depicted as curve B, where 
this quartic term vanishes, at point $D$.}
\label{special}
\end{figure}

The previous analysis assumes the existence of a generic inflection point
along the Fermi surface. This is a situation which can be achieved,
in an anisotropic system, for a finite range of values of the filling
or the chemical potential.  These points are characterized by the
absence of a quadratic term in the expansion of the dispersion relation
around the Fermi surface presented in eq. (\ref{dispersion}).
For certain values of the parameters, however, which require a fine tuning 
of the filling or the chemical potential, the cubic coefficient,
$b$, or the quartic one, $g$, in  eq. (\ref{dispersion}) can be
zero as well. Two such situations are schematically shown
in Fig.\ref{special}.

We first consider the case when the cubic 
term in the dispersion relation parallel
to the Fermi surface vanishes ($b=0$), which is realized in the $t-t'$ Hubbard
model in a square lattice (see point $G$ in the top panel of figure
\ref{special}), 
or in simple tight binding models
on the triangular lattice, for instance. 
The susceptibility becomes more anomalous than in the generic
case discussed previously, as can be seen by letting
$b\rightarrow 0$ in eq. (\ref{eq:chi-infl-interm}). One has respectively 
$A=2p_\perp-3gq_\parallel^2p_\parallel^2-2gp_\parallel^4$ 
and $B=3gq_\parallel^2$.
The condition $\alpha\gtrsim 1$ now corresponds to
$A-2k_\perp\lesssim gq_\parallel^4$, which modifies the limits of
integration in eq. (\ref{eq:parts}). 
%and the $k_\perp$ integration by parts gives
%\begin{equation}
%L=\frac{1}{\sqrt{bq_\parallel}}
%\int^{A/2}_{A/2-gq_x^4} dk_\perp \sqrt{A-2k_\perp} \Delta(vk_\perp)
%\label{parts-g}
%\end{equation}
Repeating the same arguments as before with $q_{min}=(T/gv)^{1/4}$, we 
obtain
\begin{equation}
  \label{eq:Lind-quartic}
  L= \frac{\sqrt{A/2+T/v}}{v\sqrt{g}q_\parallel}
\end{equation}
leading to
%  \chi(T)=\chi_0+\chi_1 T- \chi_3 T^{3/4}\log T
\begin{equation}
\chi =- \frac{\tilde{U}^2}{v^3} \left( \frac{T}{vg}\right)^{3/4}\log
\left( \frac{T}{g v \Lambda^4}\right)
\label{eq:chi-g0} 
\end{equation}

The other possibility is that the quartic term vanishes ($g=0$), which 
can occur in a tight-binding model with orthorombic
symmetry, considering  two different hopping parameters $t_a \neq t_b$
(see point $D$ in the bottom panel of figure
\ref{special}). In that case, however, not only $g$ but all the
even coefficients in the dispersion relation vanish. This leads to
perfect nesting between opposite branches of the Fermi surface,
giving rise to a much more singular behavior $\chi(T)\sim \log T$.

\subsection{Self-energy correction}
After integration over Matsubara frequencies, which now requires some more
attention due to the presence of two fermion lines of equal argument
(a double pole in the complex-plane integrals), the anomalous part of
the self-energy correction c) of figure \ref{fig:susc} can 
be reduced to the form
\begin{equation}
\chi \sim \tilde{U}^2 T \int\!   d^2Q d^2p d^2k
\frac{\Delta(\xi_{\mathbf{k}})\Delta(\xi_{\mathbf{p}})}
{(\Delta\xi_{\mathbf{p}}-\Delta\xi_{\mathbf{k}})^2}
\label{eq:susc-se}
\end{equation}
with  $\Delta\xi_{\mathbf{k}}=
\xi_{\mathbf{k+Q}}-\xi_{\mathbf{k}}$.
The $Q$-integration is now restricted  to the region close to
(within $T/v$ of) $\tilde{\mathbf{Q}}$. We shall not go through all the
calculations of the self-energy diagram, which can be performed
following  the same lines of
the previous sections. The results for the temperature dependence are 
analogous to those given by equations (\ref{eq:chi-parabolic}), 
(\ref{eq:chi-infl}) and (\ref{eq:chi-g0}). This can
be understood  by noting that
although the denominator in eq. (\ref{eq:susc-se}) is more singular
than the one of eq. (\ref{eq:Lind-sep}), the additional anomalies 
that it carries with it   
are cancelled by the explicit $T$ 
factor in front of eq. (\ref{eq:susc-se}), leading to the same
temperature dependence as the vertex corerction. Its sign is also the
same as the vertex diagram a).

The results of this section are summarized in table \ref{tab:results}.

%\hline
%  &  $k_\perp<\bar{k}$ & $\bar{k}<k_\perp<A/2$ \\
%\hline  $q_\parallel>0$ 
%&$\displaystyle \frac{1 \phantom{\int}}{g^{1/4}(A/2-k_\perp)^{3/4} \phantom{\int}} $ 
%&$\displaystyle \frac{-1}{(bq_\parallel)^{1/2}(A/2-k_\perp)^{1/2}}$
% \\
%\hline  $q_\parallel<0$ 
%&$\displaystyle\frac{g^{1/4} \phantom{\int}}{(bq_\parallel)(\bar{k}-k_\perp)^{1/4} \phantom{\int}}$
%&0\\ \hline

\begin{table}[htbp]
  \begin{center}
    \begin{tabular}{|l|l|}
\hline
Fermi surface geometry &  $\chi(T)$ \\
\hline
circular    & $ \chi_0+\chi_1 T$\\
inflection points (generic)  & $\chi_0+\chi_1 T- \chi_1^\prime T \log T$\\
special inflection ($b=0$) & $\chi_0+\chi_1 T- \chi_{3/4} T^{3/4}\log T$\\
nesting, saddle points     & $\chi_0 +\chi_0^\prime \log T$\\
\hline
    \end{tabular}
  \end{center}
\caption{Temperature dependence of the uniform susceptibility of an
  anisotropic 2D Fermi liquid \label{tab:results}. The linear
  contribution is always present, and is due to the  portions of the
  Fermi surface away from the inflection points. The relative
  magnitude of the regular and anomalous contributions depends on the
  degree of flatness of the Fermi surface. The special
  case $b=0$  corresponds to inflection points falling on
  particular symmetry lines of the Brillouin zone, and requires a fine
  tuning of the chemical potential (see text).}
\end{table}
\section{Examples}

%The T-dependence of the spin susceptibility in the superconducting
%cuprates is not well understood \cite{Watanabe} (and references therein).

\subsection{Superconducting cuprates}
It is often assumed that a tight-binding model on a square lattice  
with nearest neighbor ($t$)
and next-nearest neighbor ($t^\prime$) hopping reproduces well the
band structure of the layered cuprates:
\begin{equation}
\varepsilon(\mathbf{k})=-2t (\cos k_x+\cos k_y )- 4t^\prime \cos k_x
\cos k_y
\label{square}
\end{equation}
where $t^\prime / t \approx -0.25$. 
This case corresponds roughly to the top
set of Fermi surfaces in fig.\ref{special}. The dispersion relation above 
has a saddle point at a doping $\delta_{VHS}$ corresponding to a
 chemical potential is $E_{VHS} = - 4 |t^\prime|$.
The curvature of the Fermi surface along the diagonals becomes negative at
a higher doping $\delta_c$, where the chemical potential is
$E_c = -8 | t^\prime | + 16 \frac{|{t^\prime}|^3}{t^2}$.
For fillings such that $E_c \le E_F \le E_{VHS}$, the Fermi
surface has 8 inflection points. 
From these values, and the previous analysis, one can obtain a
qualitative picture of the temperature dependence of the 
susceptibilities, when the electron density is in this range:
\begin{figure}
\resizebox{8cm}{!}{\rotatebox{0}{\includegraphics{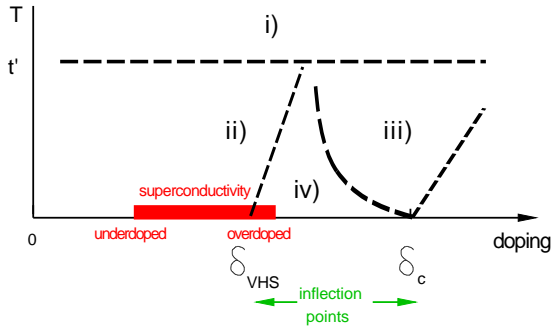}}}
\caption{Correspondence between the different shapes of
the Fermi surface discussed in the text and fillings
for the high-T$_{\rm c}$ superconductors.}
\label{VHS}
\end{figure}

%\begin{itemize}
%\item[i)]

i ) For $T \gtrsim | t^\prime |$, the susceptibility is determined
by $t$ only. As the doping is close to half filling, we 
expect $\chi ( T ) \propto | T |^0$, the result for perfect nesting.

%\item[ii)] 
ii) For $T \lesssim | t^\prime |$ and $T \gtrsim | E_F - E_{VHS} |$, 
the same behavior as in i) should be observed.

%\item[iii)] 
iii) For $T \lesssim | t^\prime |$ and $T \gtrsim | E_F - E_c |$,
the susceptibility is dominated by contribution from
the area near the special point discussed in section IIIC.
Hence, $\chi ( T ) \propto | T |^{3/4} \log T$.  

%\item[iv)] 
iv) For $T \lesssim | t^\prime |$ and $T\lesssim | E_F - E_c |$, and
$T \lesssim | E_F - E_{VHS} |$, the contributions from
the saddle point and from the special point in the previous
paragraph are absent. Thus, $\chi ( T ) \propto | T |
\log ( T )$, because of the presence of the inflection points.
%\end{itemize}

We can make the estimates of the crossover region
in the $T-$doping plane more precise from the
doping dependence of the coefficient
of the cubic term $b$ in
eq.(\ref{dispersion}).
Expanding around the saddle point, we obtain
$b \propto | E_F - E_{VHS} |$. Hence, the crossover between
regions ii) and iv) takes place at a temperature
$T^* \propto | E_F - E_{VHS} |$. Performing a similar calculation
around the situation $E_F = E_c$, we have
$b \propto \sqrt{| E_F - E_c |}$, so that
the crossover temperature is $T^* \propto ( E_c - E_F )^2$.
For fillings $E_F \sim E_c$ but with no inflection points 
in the Fermi surface, we obtain a crossover to the
$\chi ( T ) \propto | T |$ behavior due to $2 k_F$ scattering,
with a crossover temperature $T^* \propto | E_F - E_c |$.
The different regimes are schematically shown in Fig.\ref{VHS}.

Taking realistic numbers for the dispersion relation,
  our analysis predicts anomalous low
  temperature behavior in all the region between $\delta_{VHS}$ and
  $\delta_c$, corresponding to the strongly overdoped region which is
  experimentally accessible. This shows that
  non-standard behaviour of the physical properties should be expected
  even in a regime which is usually believed to be well described by
  normal Fermi liquid theory.

%\begin{figure}
%\resizebox{8cm}{!}{\rotatebox{0}{\includegraphics{phasediag.eps}}}
%\caption{Sketch of the different crossover temperatures, as function
%of density, for the band structure in eq.(\protect{\ref{square}}).}
%\label{fig:phased}
%\end{figure}

\subsection{Quasi-1D organic compounds}
\begin{figure}
\resizebox{8cm}{!}{\rotatebox{0}{\includegraphics{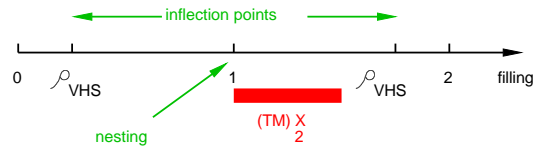}}}
\caption{Correspondence between the different shapes of
the Fermi surface discussed in the text and fillings
for compounds of the family (TM)$_2$X.}
\label{fig:org}
\end{figure}

Organic conductors are often very anisotropic due to the planar
structure of their molecules. For example, 
the salts of the family (TM)$_2$X (TM=TMTTF,TMTSF and X=inorganic
anion) are all isostructural and can be viewed as two-dimensional
arrays of weakly coupled 1D chains, since the electronic overlaps in the
transverse direction are 10 times smaller than in the chain
direction (the transfer integrals in the third direction are 500
times smaller, and can be neglected, see for instance ref.\cite{BJ99}).
The band structure is well represented as:
\begin{equation}
\varepsilon(\mathbf{k})=-2t_a \cos(k_a a )-2t_b \cos(k_b b )
\label{1d}
\end{equation}
assuming  an orthorombic structure with lattice parameters $b\approx 2a$.
This case corresponds to the bottom set of Fermi surfaces in fig.
\ref{special}.
The parameter $t_b\sim 10-30meV$ sets the scale below which the
FS is modulated in the $b$ direction, so that the predicted 
enhancement of susceptibilities
due to inflection points should be observable at and below room
temperature. The value of the anisotropy ratio $\tau=t_b/t_a$ is large enough
to ensure that the system is well described by a Fermi liquid picture
down to very low temperatures. 
The filling factor $\rho$ is fixed by charge transfer and varies from
compound to compound, ranging from $1/2$ to $1$ hole per TM site.  
%On the other hand,
%the Fermi surface shows inflection points for all fillings, so that
%the physics described here should be ubiquitous in those compounds.
The Fermi surface has two Van Hove singularities at $E_F=\pm E_{VHS}=\pm
2t(1-\tau)$, and 4 inflection points in all the interval $0<|E_F|<E_{VHS}$
Taking  $\tau= 0.1$, this corresponds to the region of fillings 
$0.3<\rho<1.7$.  
In the absence of higher harmonics in eq.(\ref{1d}),  
$E_F=0$ corresponds to half-filling ($\rho=1$), and the Fermi surface
has perfect nesting, as $\varepsilon(\mathbf{k}) =
\varepsilon(\mathbf{k} + \mathbf{Q} )$, where $\mathbf{Q} =
( \pi , \pi )$ (hopping between more distant neighbors will
suppress this effect). The points in the phase diagram where the topology of
the Fermi surface changes, leading to different 
behaviors of the electronic susceptibility, are  sketched in
figure \ref{fig:org}.

%In that case, we can expect that
%$E_F$ will not be too far from a special inflection point,
%so that $\chi ( T ) \sim | T |^{3/4}$.
  
In the (TM)$_2$X compounds,
the spin susceptibility shows a sizeable increase
in the metallic phase up to room temperature 
(see e.g. fig. 8 of reference \cite{BJ99}), which
cannot be explained by ``standard'' Fermi liquid theory (the latter predicts
variations on the scale of the Fermi temperature). On the other hand,
the presence of enhanced scattering close to inflection points
could well be the underlying mechanism of this anomalous 
temperature dependence, and should be taken into account when studying
the low temperature phase transitions of such compounds. 

Following the same procedure used in the previous subsection, 
the electron susceptibility will undergo a succession of
crossovers upon varying the filling, which can be achieved either 
by anion substitution or by  applying pressure to the samples.

%\texttt{discussion of crossovers?}

%\subsection{2D heavy fermion and other 2D materials ?}

%\section{Higher order correlation functions}

\section{Conclusions}
We have analyzed the corrections to Fermi liquid behavior
in  anisotropic interacting electronic systems in two
dimensions, which arise from the existence
of points in the Fermi surface where scattering is enhanced.
Besides the extensively studied case of
a saddle point, we have analyzed in detail the influence
of inflection points, which do not require any special fine tuning
of the chemical potential or the filling. The presence of these
points enhance the anomalous dependence on
temperature which arise from
$2 k_F$ scattering in isotropic Fermi surfaces\cite{CM01}.
We find that the corrections which were linear in $|T| $ change
into $ |T|  \log  |T|$. The absence of symmetries also implies
the lack of cancellation between different diagrams, so that
these anomalies should be observed in both the spin and charge
susceptibilities.

For special fillings, more
singular behavior is expected. In the case of systems with
tetragonal or hexagonal symmetry, when the Fermi surface
is close to these fillings, the corrections to the susceptibilities
go as $ |T|^{3/4} \log  |T|  $, showing that
the existence of non integer $T$ dependences does not
need to violate Landau's model for the low energy
excitations of a Fermi liquid.   

We have also discussed the possible crossovers between the
different regimes analyzed, and the experimental consequences
that they may lead to. In the case of the superconducting cuprates,
anomalous susceptibilities should appear in the strongly overdoped
region, above the doping $\delta_{VHS}$ characterized by Van Hove
singularities in the density of states. On the other hand, all of  the
organic conductors of the family (TM)$_2$X should fall in the region
of fillings where anomalous corrections to the susceptibility are important. 
Of course, the analysis presented here should also apply to other
classes of quasi two-dimensional systems (heavy fermion 
materials, Sr$_2$RuO$_4$, electrically doped 2D organic
films, other organic conductors \ldots).

%It is worth noting that 
%although we focused here explicitely on the 
%\textit{spin} susceptibility, the \textit{charge} response should exhibit the
%same  behavior summarized in table I. 
%Indeed, one can argue that the cancellations 
%occurring in the charge sector, leading to a charge susceptibility
%$\chi_{ch}\sim T^2$ instead of $T$, as pointed out in reference
%\cite{CM01}, are specific to the case of a
%perfectly isotropic FS, and are not effective in general.

Finally, let us point out that the breakdown of the 
Sommerfeld expansion for the spin susceptibility suggests that
the free energy ${\cal F}$ itself has a non analytic dependence
on $T$, once that high order interactions ($2k_F$ scattering) 
are taken into account.  If,  as was proposeded 
in\cite{M88,M99,M01}, and numerically verified in \cite{HK99}, 
 the role of temperature  and magnetic field is interchangeable in the
functional form  of ${\cal  F}(T,H)$, one can conclude that
the anomalous $T$-dependences calculated here for the susceptibility are 
also expected in the specific-heat coefficient $\gamma=C/T$.

\section{Acknowledgements}
We are thankful to R. Markiewicz and M. A. H. Vozmediano for helpful
discussions. This work was financially supported by MEC (Spain)
through grant PB96-0875, and the European Union through
grant FMRXCT980183.
\appendix

%\section{T-dependence with inflection points}
%integration can be divided into different regions, according to 
%the sign and magnitude of $a$, which reflect in the conditions of
%table \ref{tab:regions} (see appendix \ref{ap:infl}). 
%We shall focus on the region close to the $k_\perp$-edge at
%%There are three different regions of phase space contributing to the
%%susceptibility, and which are summarized in table 1
%\label{ap:infl}
%\begin{table}[htbp]
%  \begin{center}
%    \begin{tabular}{|c|c|c|}
%\hline
%  &  $k_\perp<\bar{k}$ & $\bar{k}<k_\perp<A/2$ \\
%\hline  $q_\parallel>0$ 
%&$\displaystyle \frac{1 \phantom{\int}}{g^{1/4}(A/2-k_\perp)^{3/4} \phantom{\int}} $ 
%&$\displaystyle \frac{-1}{(bq_\parallel)^{1/2}(A/2-k_\perp)^{1/2}}$
% \\
%\hline  $q_\parallel<0$ 
%&$\displaystyle\frac{g^{1/4} \phantom{\int}}{(bq_\parallel)(\bar{k}-k_\perp)^{1/4} \phantom{\int}}$
%&0\\ \hline
%    \end{tabular}
%\caption{The function to be integrated in different 
%regions of momentum space. It is $0$ for $k_\perp>A/2$, and we have defined
%$\bar{k}=A/2-(9/16)b^2q_\parallel^2/g$.}
%    \label{tab:regions}
%  \end{center}
%\end{table}

\section{Inflection points in the $t-t^\prime$ model}

%\paragraph*{Expansion around critical doping}
We shall determine here the parameters of the dispersion relation 
(\ref{dispersion})  in the case of a tight-binding model on a square
lattice with nearest
($t$) and next-nearest ($t^\prime$) neighbor hopping.  
%in order to evaluate the prefactor of the anomalous correction to
%the linear susceptibility. 
Let us focus to the doping levels
close to $\delta_c$, the point where the inflection points of the
Fermi surface merge in
pairs on the diagonals of the BZ, leading to the most singular
corrections to the susceptibility.
It is then natural to
rewrite the dispersion relation in a basis rotated by $45^o$:
\begin{equation}
\xi=-4t (\cos p \cos q )+4t^\prime (\cos^2 p-\sin^2 q)-E_F
\end{equation}
where $p=(k_x+k_y)/2$ and  $q=(k_x-k_y)/2$. 
%(this is not a pure rotation, but the formulae are greatly simplified). 
The Fermi surface crosses
the diagonal ($q=0$) at a momentum $p_F$ given by
$  E_F=-4t \cos p_F+4t^\prime \cos^2 p_F$.
%\begin{equation}
%  \label{eq:Ef}
%\end{equation}
The dispersion relation can then be expanded  as
\begin{equation}
  \label{eq:disp-exp}
\xi = A (p-p_F) +Bq^2+Cq^4
\end{equation}
with $A=4[t \sin p_F - t^\prime \sin 2p_F]$,  
$B=2t\cos p_F -4 t^\prime$ and
$C= (4/3) [t^\prime - (t/8)\cos p_F]$.
The topology of the Fermi surface changes at two well-defined doping levels: 
\begin{itemize}
\item
the curvature changes sign at a doping 
$\delta=\delta_c$  given by the condition  $B=0$. The
corresponding Fermi energy is $E_c=-8  t^\prime +16 (t^\prime)^3/t^2$  
and the coordinates of the inflection point are 
$(p_c,q_c)=(\arccos 2t^\prime/t,0)$, corresponding to the point 
$G$ of figure \ref{special}; 
%When $t^\prime=0$, i.e. in the simplest tight-binding model, we
%recover $E_c=0$ indicating that the most singular
%situation occurs at half filling. 
\item
Van Hove singularities arise at a doping
$\delta_{VHS}$ given by $E_{VHS}=-4 t^\prime$ (M-points in figure 
 \ref{special}). 
\end{itemize}
Inflection points appear in all the region of dopings 
$\delta_{VHS}<\delta<\delta_c$, following the dashed curve of figure 
\ref{special} (top panel). The consequences on the physical properties of the
system are summarized in figure \ref{VHS}.

\paragraph*{Dispersion around inflection points.}
The equation of the Fermi surface is $\xi=0$, which implicitly defines
a function $p=p(q)$. Putting
the second derivative $p^{\prime\prime}(q)=0$ yields the coordinates
$(p_0,q_0)$ of the inflection points. For $\delta\approx\delta_c$, setting 
 $u=1-4 (t^\prime/t)^2$, we can write
\begin{equation}
p_0=p_c+\frac{E_F-E_c}{4tu^{3/2}}\;\; ; \;\;
q_0=\left(\frac{E_F-E_c}{12 u t^\prime}\right)^{1/2}
\end{equation}
so that the trajectory of the inflection points is parabolic around $G$.
%, their separation between the two symmetric inflection points varies 
%as $(E_F-E_c)^{1/2} \sim (\delta_c-\delta)^{1/2}$. 
By expanding around $(p_0,q_0)$, we obtain an equation of the form
(\ref{dispersion}) with 
\begin{equation}
 v=4 t u^{3/2}\; ; \;
  b=\frac{t^\prime}{t}\left(\frac{E_F-E_c}{12
      t^\prime}\right)^{1/2}\; ;\;
  g=\frac{t^\prime}{4 t u^{3/2}}
\end{equation}
%\begin{eqnarray}
%  v&=&4 t u^{3/2}\\
%  b&=&\frac{t^\prime}{t}\left(\frac{E_F-E_c}{12
%      t^\prime}\right)^{1/2}\\
%  g&=&\frac{t^\prime}{4 t u^{3/2}}
%\end{eqnarray}

% Taking $N(E_F)\sim N(E_c)\sim 1/t$, 
%the energy scale of the logarithmic correction is
%\begin{equation}
%  \label{eq:Wscale}
%  W_{t-t^\prime}=
%  \frac{\beta v^3}{t^3}\sim 2^6 u^{9/2}\sqrt{t^\prime(E_F-E_c)}
%\end{equation}

\section{Inflection points in the $t_a-t_b$ model}

We shall now derive the parameters of eq. (\ref{dispersion}) for
a tight binding model on an orthorombic lattice, with anisotropic hopping
($\tau=t_b/t_a\ll 1$). Let us rewrite for simplicity the dispersion
relation (\ref{1d}) as
\begin{equation}
\xi=-2t_a [\cos k +\tau \cos p +\nu]
\label{1dbis}
\end{equation}
with $k=k_a a$, $p=k_b b$ and $\nu=E_F/2t$. The equation of
the Fermi surface is $k=\arccos(-\nu-\tau \cos p)$. The number of
electrons per site  is approximately given by 
$\rho=2\pi^{-1}\arccos(-\nu)$. The Fermi surface has 2
Van Hove singularities at $E_{VHS}=\pm 2t (1-\tau)$, and 4 inflection points
for any $0<|E_F|<E_{VHS}$.  At half filling ($E_F=0$), the two
branches of the open Fermi surface are perfectly nested. 
The physical consequences of the changes in
the Fermi surface topology occurring at those special fillings 
are sketched in figure \ref{fig:org}.

\paragraph*{Dispersion around inflection points.}
By setting 
$k^{\prime\prime}(p)=0$ we find that the
inflection points are located at
\begin{equation}
p_0=\arccos\left( \frac{\tau\nu}{1-\nu^2}\right) \;\; ; \;\;
k_0=\arccos\left( -\nu- \frac{\tau^2\nu}{1-\nu^2}\right)
\end{equation}
For filling levels close to $\nu=0$ (half filling), the location of
the inflection points describes a
straight line of slope $-\tau/\sqrt{1-\nu^2}$ in the $(k,p)$
plane (see bottom panel of figure \ref{special}). After a rotation  of the coordinate axes, 
we obtain an equation of the form (\ref{dispersion}) with
\begin{equation}
v=2 t \sqrt{1-\nu^2}\; ; \; b=\frac{\tau}{6\sqrt{1-\nu^2}}\; ; \;
 g=-\frac{\nu\tau^2}{24(1-\nu^2)^{3/2}}
\end{equation}
The approach to the perfect nesting situation at half filling is
signalled by a vanishing $g$, the coefficient of the quartic term in the
dispersion relation.

%\end{multicols}

\end{document}